\documentclass[10pt]{article}
\usepackage{graphicx}
\usepackage{amssymb}
\usepackage{amsfonts}
\usepackage{amsmath}
\usepackage{amsthm}

\numberwithin{equation}{section}

\newtheorem*{acknowledgement}{Acknowledgement}

\begin{document}

\title{Simple model of bouncing ball dynamics. Displacement of the limiter
assumed as a cubic function of time.}
\author{Andrzej Okni\'nski$^{1)}$, Bogus{\l }aw Radziszewski$^{2)}$ \\
Kielce University of Technology, 25-314 Kielce, Poland$^{1)}$\\
Collegium Mazovia Innovative University, 08-110 Siedlce, Poland$^{2)}$\\
}
\maketitle

\begin{abstract}
Nonlinear dynamics of a bouncing ball moving vertically in a gravitational
field and colliding with a moving limiter is considered and the Poincar\'{e}
map, describing evolution from an impact to the next impact, is described.
Displacement of the limiter is assumed as periodic, cubic function of time.
Due to simplicity of this function analytical computations are possible.
Several dynamical modes, such as fixed points, $2$ - cycles and chaotic
bands are studied analytically and numerically. It is shown that chaotic
bands are created from fixed points after first period doubling in a
corner-type bifurcation. Equation for the time of the next impact is solved
exactly for the case of two subsequent impacts occurring in the same
period of limiter's motion making analysis of chattering possible.
\end{abstract}

\section{Introduction}

In the present paper we study dynamics of a small ball moving vertically in
a gravitational field and impacting with a periodically moving limiter (a
table). This model belongs to the field of nonsmooth and nonlinear dynamical
systems \cite{diBernardo2008,Luo2006,Awrejcewicz2003,Filippov1988}. In such
systems nonstandard bifurcations such as border-collisions and grazing
impacts leading often to complex chaotic motions are typically present. It
is important that nonsmooth systems have many applications in technology 
\cite{Stronge2000,Mehta1994,Knudsen1992,Wiercigroch2008}.

In the bouncing ball dynamics it is usually difficult or even impossible to
solve nonlinear equation for an instant of the next impact. We approached
this problem assuming a special motion of the table. Recently, we have
considered several models of motion of a material point in a gravitational
field colliding with a limiter moving periodically with piecewise constant
velocity \cite{AOBR2009,AOBR2010a} and velocity depending linearly on time 
\cite{AOBR2010b}. In the present work we study the model in which periodic
displacement of the table is a cubic function of time, carrying out our
project to approximate the sinusoidal motion of the table as exactly as
possible but preserving possibility of analytical computations \cite%
{AOBR2010c}.

The paper is organized as follows. In Section 2 a one dimensional dynamics
of a ball moving in a gravitational field and colliding with a table is
reviewed and the corresponding Poincar\'{e} map is constructed. A
bifurcation diagram is computed for displacement of the table assumed as
cubic and periodic function of time. In the next Section dynamical modes
shown in the bifurcation diagram such as fixed points, $2$ - cycles and
chaotic bands as well as the case of $N$ impacts in one interval of the
limiter's motion are studied analytically and numerically. We summarize our
results in Section 4.

\section{Bouncing ball: a simple motion of the table}

Let a ball moves vertically in a constant gravitational field and collides
with a periodically moving table. We treat the ball as a material point and
assume that the limiter's mass is so large that its motion is not affected
at impacts. Dynamics of the ball from an impact to the next impact can be
described by the following Poincar\'{e} map in nondimensional form \cite%
{AOBR2007} (see also Ref. \cite{Luo1996} where analogous map was derived
earlier and Ref. \cite{Luo2009a} for generalizations of the bouncing ball
model):
\begin{subequations}
\label{TV}
\begin{eqnarray}
\gamma Y\left( T_{i+1}\right)  &=&\gamma Y\left( T_{i}\right) -\Delta
_{i+1}^{2}+\Delta _{i+1}V_{i},  \label{T} \\
V_{i+1} &=&-RV_{i}+2R\Delta _{i+1}+\gamma \left( 1+R\right) \dot{Y}\left(
T_{i+1}\right) ,  \label{V}
\end{eqnarray}%
where $T_{i}$ denotes time of the $i$-th impact and $V_{i}$ is the
corresponding post-impact velocity while $\Delta _{i+1}\equiv T_{i+1}-T_{i}$%
. The parameters $\gamma $, $R$ are a nondimensional acceleration and the
coefficient of restitution, $0\leq R<1$ \cite{Stronge2000}, respectively and
the function $Y\left( T\right) $ represents the limiter's motion. 

The table's motion has been usually assumed in form $Y_{s}(T)=\sin (T)$, cf. 
\cite{Luo1996,Luo2009a} and references therein. In this case it is basically
impossible to solve the Eq.(\ref{T}) for $T_{i+1}$. Accordingly, we have
decided to choose the limiter's periodic motion in a polynomial form to make
analytical investigations of the dynamics possible. In our previous papers
we have assumed displacement of the table as piecewise linear periodic
function of time \cite{AOBR2009,AOBR2010a} as well as quadratic \cite%
{AOBR2010b}. In this work we study dynamics for a cubic function of time $%
Y_{c}\left( T\right) $: 
\end{subequations}
\begin{subequations}
\label{C1}
\begin{eqnarray}
Y_{c}\left( T\right)  &=&12\sqrt{3}\hat{T}\left( \hat{T}-\tfrac{1}{2}\right)
\left( \hat{T}-1\right) ,  \label{Ci} \\
\dot{Y}_{c}\left( T\right)  &=&6\sqrt{3}\left( 6\hat{T}^{2}-6\hat{T}%
+1\right) ,  \label{Cii}
\end{eqnarray}%
with $\hat{T}=T-\left\lfloor T\right\rfloor $, where $\left\lfloor
x\right\rfloor $ is the floor function -- the largest integer less than or
equal to $x$ (i.e. $0\leq \hat{T}\leq 1$).

Since the period of motion of the limiter is equal to one, the map (\ref{TV}%
) is invariant under the translation $T_{i}\rightarrow T_{i}+1$.
Accordingly, all impact times $T_{i}$ can be reduced to the unit interval $%
\left[ 0,\ 1\right] $. The model consists thus of equations (\ref{TV}), (\ref%
{C1}) with control parameters $R$, $\gamma $.

\begin{figure}[ht!]
\center
\includegraphics[scale=0.7]{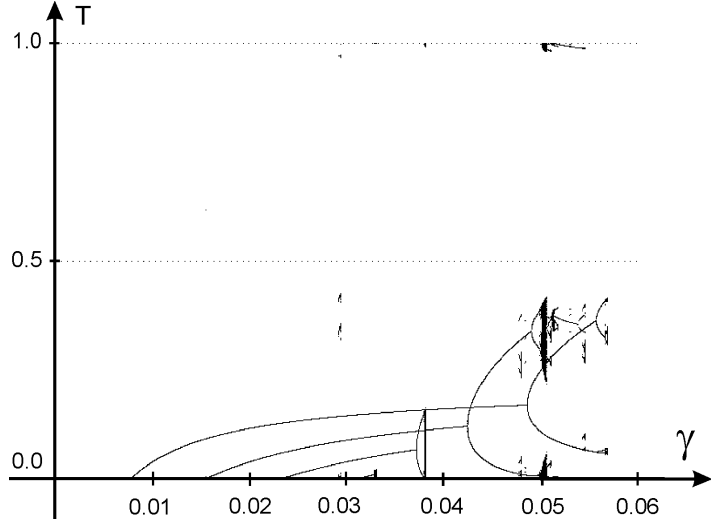}
\caption{Bifurcation diagram, $R = 0.85$, $\protect\gamma \in [0,\ 0.06]$.}
\label{F1}
\end{figure}

In Fig. \ref{F1} above we show the bifurcation diagram with impact times
versus $\gamma $ computed for growing $\gamma $ and $R=0.85$. It follows
that dynamical system (\ref{TV}), (\ref{C1}) has several attractors: two
fixed points which after one period doubling give rise to chaotic bands and
two other fixed points which go to chaos via period doubling scenario. There
are also several small attractors. We shall investigate some of these
attractors in the next Section combining analytical and numerical approach.
General analytical conditions for birth of new modes of motion were given in 
\cite{Luo2009b}.

\section{Analytical and numerical results}

\subsection{Fixed points and their stability}

We shall first study periodic solutions with one impact per $k$ periods.
Such states have to fulfill the following conditions: 
\end{subequations}
\begin{equation}
V_{n+1}=V_{n}\equiv V_{\ast }^{\left( k/1\right) },\ T_{n+1}=T_{n}+k\equiv
T_{\ast }^{\left( k/1\right) }+k\qquad \left( k=1,2,\ldots \right) ,
\label{condk1a}
\end{equation}%
where:%
\begin{equation}
T_{\ast }^{\left( k/1\right) }\in \left( 0,\ 1\right) ,\ V_{\ast }^{\left(
k/1\right) }>\gamma \dot{Y}_{c_{1}}\left( T_{\ast }^{\left( k/1\right)
}\right) .  \label{condk1b}
\end{equation}

Substituting these conditions into (\ref{TV}), (\ref{C1}) we obtain two sets
of fixed points:%
\begin{eqnarray}
0 &\leq &T_{\ast \left( s\right) }^{\left( k/1\right) }=\tfrac{1}{2}-\tfrac{%
\sqrt{3}}{18\gamma }\sqrt{9\gamma ^{2}+\sqrt{3}k\gamma \tfrac{1-R}{1+R}}\leq 
\tfrac{1}{2}  \label{FixStA} \\
V_{\ast }^{\left( k/1\right) } &=&k  \notag
\end{eqnarray}%
where the impact occurs in time interval $T_{\ast \left( s\right) }^{\left(
k/1\right) }\in \left( 0,\ \frac{1}{2}\right) $ and

\begin{eqnarray}
\tfrac{1}{2} &\leq &T_{\ast \left( u\right) }^{\left( k/1\right) }=\tfrac{1}{%
2}+\tfrac{\sqrt{3}}{18\gamma }\sqrt{9\gamma ^{2}+\sqrt{3}k\gamma \tfrac{1-R}{%
1+R}}\leq 1  \label{FixUStA} \\
V_{\ast }^{\left( k/1\right) } &=&k  \notag
\end{eqnarray}%
with impacts taking place in time interval $T_{\ast \left( u\right)
}^{\left( k/1\right) }\in \left( \frac{1}{2},\ 1\right) $.

Solutions (\ref{FixStA}) fulfill physical requirements and are stable in the
following interval of $\gamma $:%
\begin{equation}
\tfrac{\sqrt{3}}{18}k\tfrac{1-R}{1+R}\leq \gamma \leq \tfrac{\sqrt{3}}{%
54(1+R)^{2}}\left( 3k(R^{2}-1)+\sqrt{9k^{2}(R^{2}-1)^{2}+12(R^{2}+1)^{2}}%
\right)  \label{exst1}
\end{equation}%
where lower bound is a consequence of $T_{\ast \left( s\right) }^{\left(
k/1\right) }\geq 0$ while the upper bound follows from the condition that
eigenvalues $\lambda $ of the stability matrix obey $\left\vert \lambda
\right\vert <1$. Accordingly, for $R=0.85$ there are only four stable fixed
points with $k=1, \ 2, \ 3, \ 4$, shown in Fig. \ref{F1} -- they appear at $%
T = 0$ and $\gamma = 0.0078, \ 0.0156, \ 0.0234, \ 0.0312$, respectively.

On the other hand, solutions (\ref{FixUStA}) are always unstable and are
physical for:%
\begin{equation}
\tfrac{\sqrt{3}}{18}k\tfrac{1-R}{1+R}\leq \gamma ,  \label{exust1}
\end{equation}%
what is equivalent to the condition $T_{\ast \left( u\right) }^{\left(
k/1\right) }\leq 1$.

\subsection{Birth of stable $2$ - cycles and transition to chaos}

Let us note that $2$ - cycles are created from fixed points such that $%
T_{i}\in \left[ 0,1\right] ,\ T_{i+1}\in \left[ k,k+1\right] $ with $%
k=1,2,3,4$. Therefore equations defining $2$ - cycles read:%
\begin{equation}
\left\{ 
\begin{array}{l}
\gamma Y\left( \hat{T}_{i+1}\right) =\gamma Y\left( T_{i}\right) -\left(
T_{i+1}-T_{i}\right) ^{2}+\left( T_{i+1}-T_{i}\right) V_{i} \\ 
V_{i+1}=-RV_{i}+2R\left( T_{i+1}-T_{i}\right) +\gamma \left( 1+R\right) \dot{%
Y}\left( \hat{T}_{i+1}\right) \\ 
\hat{T}_{i+1}\equiv T_{i+1}-k \\ 
\gamma Y\left( \hat{T}_{i+2}\right) =\gamma Y\left( \hat{T}_{i+1}\right)
-\left( T_{i+2}-\hat{T}_{i+1}\right) ^{2}+\left( T_{i+2}-\hat{T}%
_{i+1}\right) V_{i+1} \\ 
V_{i+2}=-RV_{i+1}+2R\left( T_{i+2}-\hat{T}_{i+1}\right) +\gamma \left(
1+R\right) \dot{Y}\left( \hat{T}_{i+2}\right) \\ 
\hat{T}_{i+2}\equiv T_{i+2}-k \\ 
\hat{T}_{i+2}=T_{i} \\ 
V_{i+2}=V_{i}%
\end{array}%
\right.  \label{2cycleG}
\end{equation}%
where $k=1,2,3,4$ and $Y,\ \dot{Y}$ are given by (\ref{C1}). We know that $2$
- cycles appear for $\gamma =\tfrac{\sqrt{3}}{54(1+R)^{2}}\left( 3k(R^{2}-1)+%
\sqrt{9k^{2}(R^{2}-1)^{2}+12(R^{2}+1)^{2}}\right) $, cf. Eq.(\ref{exst1}).
We have solved the set of equations (\ref{2cycleG}) in closed form but we
skip these lengthy formulae because of lack of space.

The bifurcation diagram shown in Fig. 1 suggests that in the case of two
fixed points with $k=3,4$ transition to chaos occurs after the first period
doubling when $T_{\ast 1}=0$. Therefore, in order to determine values of
parameters at which the transition to chaos occurs, we have to solve Eq.(\ref%
{2cycleG}) with condition $T_{i}=0$. After this substitution equations (\ref%
{2cycleG}) are easily solved to yield:

\begin{equation}
\left\{ 
\begin{array}{l}
a_{4}X^{4}+a_{3}X^{3}+a_{2}X^{2}+a_{1}X+a_{0}=0\smallskip \\ 
a_{4}=\left( R+1\right) ^{2} \\ 
a_{3}=-k\left( R+1\right) \left( 7R+1\right) \\ 
a_{2}=k\left( 10k+3\right) R^{2}+\left( 10R-8\right) k^{2}-3k \\ 
a_{1}=4k^{2}\left( \left( 3-R\right) k+3\right) \\ 
a_{0}=-2k^{2}\left( k+1\right) \left( 2k+1\right) \left( 1+R^{2}\right)
\smallskip \\ 
\gamma _{cr}^{\left( k\right) }=\frac{\sqrt{3}X}{18\left( 1+R\right) }\frac{%
-\left( R+1\right) ^{2}X+4Rk}{\left( 2+4R\right) X^{3}-3\left( R+1\right)
\left( 2k+1\right) X^{2}+6k\left( k+1\right) X+k\left( k+1\right) \left(
2k+1\right) \left( R-1\right) }%
\end{array}%
\right.  \label{sol}
\end{equation}%
where $X\equiv T_{i+1}$.

For example, for $R=0.85$ and $k=4$ we get from (\ref{sol}) $\gamma
_{cr}^{\left( 4\right) }\cong 0.03284$ while for $R=0.85$ and $k=3$ we
obtain $\gamma _{cr}^{\left( 3\right) }\cong 0.03806$ and indeed, precisely
at this point on the $\gamma $ axis, branches of the corresponding $2$ -
cycles reach values $T_{\ast 1}=0$ and transform into chaotic bands. This
scenario does not apply for $R=0.85$ and $k=1,2$ - in the case $k=2$ there
is another period doubling prior to $\gamma _{cr}^{\left( 2\right) }$ while
for $k=1$ physical solutions of (\ref{sol}) do not exist.

\subsection{$N$ impacts in one period of limiter's motion and chattering}

In the bouncing ball dynamics chattering and chaotic dynamics arise
typically, see \cite{Giusepponi2003,Giusepponi2005} where chattering
mechanism was studied numerically for sinusoidal motion of the table. Due to
simplicity of our model analytical computations are possible.

Let us assume that $N=2$ impacts, $T_{i}$, $T_{i+1}$, occur in the same
period. Then the solution $\Delta _{i+1}=0$ of equation (\ref{T}) is always
present. We thus obtain from Eqs.(\ref{T}) and (\ref{Ci}): 
\begin{subequations}
\label{DD}
\begin{eqnarray}
\Delta _{i+1} &=&0,\quad \frac{1}{4G}\left( 3G-2-6GT_{i}\pm \sqrt{D}\right) ,
\label{Delta} \\
D &=&-12G^{2}T_{i}^{2}+\left( 24G+12G^{2}\right) T_{i}+G^{2}-12G+16GV_{i}+4,
\label{D}
\end{eqnarray}
where $G\equiv 12\sqrt{3}\gamma $, $\Delta _{i+1}\equiv T_{i+1}-T_{i}$. We
can now rewrite Eqs.(\ref{TV}) in simplified form: 
\end{subequations}
\begin{subequations}
\label{EQS}
\begin{eqnarray}
&&\hspace{-51pt}\left\{ 
\begin{array}{l}
T_{i+1}=T_{i}+\Delta _{i+1} \\ 
v_{i+1}=-Rv_{i}+2R\Delta _{i+1}+3RG\Delta _{i+1}\left( \Delta
_{i+1}+2T_{i}-1\right)%
\end{array}
\right.  \label{eqs} \\
\Delta _{i+1} &=&\tfrac{1}{4G}\left( 3G-2-6GT_{i}+\sqrt{\left(
3G-2-6GT_{i}\right) ^{2}+16Gv_{i}}\right)  \label{del}
\end{eqnarray}
where $v_{i}=V_{i}-G\left( 3T_{i}^{2}-3T_{i}+\tfrac{1}{2}\right) $ is a
relative velocity and the solution $\Delta _{i+1}>0$, cf. Eqs. (\ref{DD}),
was chosen. Equations (\ref{EQS}) define an \emph{explicit} nonlinear map.
This formalism makes possible analysis of chattering and grazing in the $%
N\rightarrow \infty $ limit. It follows immediately that grazing manifold, $%
v_{\ast }=0,\ T_{i+1}=T_{i}\equiv T_{\ast }$, i.e. $\Delta _{i+1}=0$, exists
only for $3G-2-6GT_{\ast }\leq 0$ and hence for 
\end{subequations}
\begin{equation}
1\geq T_{\ast }\geq \max \left( T_{cr},\ 0\right) ,\quad T_{cr}\overset{df}{%
= }\tfrac{1}{2}-\tfrac{1}{3G}.  \label{manifold}
\end{equation}

After computing eigenvalues $\Lambda _{1,2}$ of the stability matrix $S$\ on
the grazing manifold, $v_{\ast }=0,\ 3G-2-6GT_{\ast }\leq 0$, we get after
straightforward calculations $\Lambda _{1}=1$, $\Lambda _{2}=R<1$. It thus
follows that the grazing manifold is always attracting in one eigendirection
and neutral in another ($T$).

In the final stage of grazing $v_{i}$ is very small. Let $3G-2-6GT_{i}<0$.
It follows that for $0<x_{i}<1$, $x_{i}\overset{df}{=}\tfrac{16Gv_{i}}{%
\left( -3G+2+6GT_{i}\right) ^{2}}$, we get the convergent expansion of the
square-root in (\ref{del}): 
\begin{equation}
\Delta _{i}=\tfrac{-3G+2+6GT_{i}}{4G}\left( \tfrac{1}{2}x_{i}-\tfrac{1}{8}%
x_{i}^{2}+\ldots \right) \cong \tfrac{2v_{i}}{-3G+2+6GT_{i}}
\label{approxDelta}
\end{equation}%
and the inequality $x_{i}<1$ implies:%
\begin{equation}
T_{i}>T_{A_{1}}\equiv \left( \tfrac{1}{2}-\tfrac{1}{3G}\right) +\tfrac{2}{3}%
\sqrt{\tfrac{v_{i}}{G}}.  \label{TA1}
\end{equation}

Approximate equations describing chattering, obtained from the approximate
equation (\ref{approxDelta}) and exact equations (\ref{eqs}), are of form: 
\begin{subequations}
\label{EQSapprox}
\begin{eqnarray}
T_{i+1} &=&T_{i}+\tfrac{2v_{i}}{-3G+2+6GT_{i}},  \label{approxT} \\
v_{i+1} &=&\lambda _{i}v_{i},\quad \lambda _{i}\overset{df}{=}R\left( 1+%
\tfrac{12Gv_{i}}{\left( -3G+2+6GT_{i}\right) ^{2}}\right) ,  \label{approxV}
\end{eqnarray}%
and to ensure convergence $v_{i}\rightarrow 0$ we assume that $\lambda
_{i}<1 $ (i.e. $x_i<\frac{4}{3}(R^{-1}-1)$) -- this leads to the following
condition for $T_{i}$: 
\end{subequations}
\begin{equation}
T_{i}>T_{A_2}\equiv \left( \tfrac{1}{2}-\tfrac{1}{3G}\right) +\sqrt{\tfrac{
Rv_{i}}{3\left( 1-R\right) G}}.  \label{TA2}
\end{equation}

It is now possible to estimate time of $N$-th impact:%
\begin{equation}
T_{\left( N\right) }\overset{df}{=}T_{i}+\sum\nolimits_{j=i+1}^{i+N}\Delta
_{j}.  \label{TN1}
\end{equation}%
The time $T_{\left( N\right) }$ can be computed exactly since%
\begin{equation}
\Delta _{j+1}=\tfrac{2v_{j}}{-3G+2+6GT_{j}}=\tfrac{2v_{j-1}\lambda _{j-1}}{%
-3G+2+6G\left( T_{j-1}+\Delta _{j}\right) }=\tfrac{2v_{j-1}}{-3G+2+6GT_{j-1}}%
R=R\Delta _{j},  \label{NewDelta}
\end{equation}%
where equations (\ref{approxDelta}), (\ref{approxV}) were used, and thus%
\begin{equation}
T_{\left( N\right) }=T_{i}+\tfrac{2v_{i}}{-3G+2+6GT_{i}}\sum%
\nolimits_{j=1}^{N}R^{j-1}=T_{i}+\tfrac{2v_{i}}{-3G+2+6GT_{i}}\tfrac{1-R^{N}%
}{1-R}.  \label{TN2}
\end{equation}

If the ball impacts at time $T_{cr}<T_{i}<1$, performs infinite number of
impacts and sticks (in one interval of limiter's motion) then the time is $%
T_{\left( \infty \right) }$. Solving the inequality $T_{\left( \infty
\right) }<1$ we get conditions for grazing: 
\begin{eqnarray}
T_{i} &>&T_{B}\equiv \tfrac{1}{12G}\left( \left( -2+9G\right) -\sqrt{\left(
3G+2\right) ^{2}-\tfrac{48Gv_{i}}{1-R}}\right) >T_{cr}\,,  \label{TB} \\
T_{i} &<&T_{C}\equiv \tfrac{1}{12G}\left( \left( -2+9G\right) +\sqrt{\left(
3G+2\right) ^{2}-\tfrac{48Gv_{i}}{1-R}}\right) <1\,.  \label{TC}
\end{eqnarray}

Finally, the (approximate) condition for grazing is: $\max \left(
T_{A_{1},}T_{A_{2}},T_{B}\right) <T_{i}<T_{C}$, cf. Eqs. (\ref{TA1}), ( \ref%
{TA2}), (\ref{TB}), (\ref{TC}).

\section{Summary}

It has been shown that some chaotic bands are created from fixed points
after first period doubling in a corner-type bifurcation and critical values
of control parameter $\gamma $ have been determined. Equations for $N$
impacts in one period of limiter's motion were found and simplified
significantly, making analysis of chattering and grazing possible.
Approximate equations describing final stage of chattering were obtained and
(approximate) condition for grazing was computed in analytical form.

\begin{acknowledgement}
The paper was presented at the 11th Conference on Dynamical Systems --
Theory and Applications. December 5-8, 2011. \L \'{o}d\'{z}, POLAND.
\end{acknowledgement}


\begin{thebibliography}{99}
\bibitem{diBernardo2008} M. di Bernardo, C.J. Budd, A.R. Champneys, P.
Kowalczyk, \textit{Piecewise-Smooth Dynamical Systems. Theory and
Applications}. Series: Applied Mathematical Sciences, vol. 163. Springer,
Berlin (2008).

\bibitem{Luo2006} A.C.J.Luo, \textit{Singularity and Dynamics on
Discontinuous Vector Fields}. Monograph Series on Nonlinear Science and
Complexity, vol. 3. Elsevier, Amsterdam (2006).

\bibitem{Awrejcewicz2003} J. Awrejcewicz, C.-H. Lamarque, \textit{%
Bifurcation and Chaos in Nonsmooth Mechanical Systems}.World Scientific
Series on Nonlinear Science: Series A, vol. 45. World Scientific Publishing,
Singapore (2003).

\bibitem{Filippov1988} A.F. Filippov, \textit{Differential Equations with
Discontinuous Right-Hand Sides}. Kluwer Academic, Dordrecht (1988).

\bibitem{Stronge2000} W.J. Stronge, \textit{Impact mechanics}. Cambridge
University Press, Cambridge (2000).

\bibitem{Mehta1994} A. Mehta (ed.), \textit{Granular Matter: An
Interdisciplinary Approach}. Springer, Berlin (1994).

\bibitem{Knudsen1992} C. Knudsen, R. Feldberg, H. True, Bifurcations and
chaos in a model of a rolling wheel-set. Philos. Trans. R. Soc. Lond. A 
\textbf{338}, 455--469 (1992).

\bibitem{Wiercigroch2008} M. Wiercigroch, A.M. Krivtsov, J. Wojewoda,
Vibrational energy transfer via modulated impacts for percussive drilling,
Journal of Theoretical and Applied Mechanics \textbf{46}, 715--726 (2008).

\bibitem{AOBR2009} A. Okni\'nski, B. Radziszewski, Dynamics of impacts with
a table moving with piecewise constant velocity, Nonlinear Dynamics \textbf{%
58}, 515--523 (2009).

\bibitem{AOBR2010a} A. Okni\'nski, B. Radziszewski, Chaotic dynamics in a
simple bouncing ball model, Acta Mech. Sinica \textbf{27}, 130--134 (2011),
arXiv:1002.2448 [nlin.CD] (2010).

\bibitem{AOBR2010b} A. Okni\'nski, B. Radziszewski, Simple model of bouncing
ball dynamics: displacement of the table assumed as quadratic function of
time, Nonlinear Dynamics \textbf{67}, 1115---1122 (2012).

\bibitem{AOBR2010c} A. Okni\'nski, B. Radziszewski, Simple models of
bouncing ball dynamics and their comparison, arXiv:1002.2448 [nlin.CD]
(2010).

\bibitem{AOBR2007} A. Okni\'nski, B. Radziszewski, Grazing dynamics and
dependence on initial conditions in certain systems with impacts,
arXiv:0706.0257 [nlin.CD] (2007).

\bibitem{Luo1996} A.C.J. Luo, R.P.S. Han, The dynamics of a bouncing ball
with a sinusoidally vibrating table revisited, Nonlinear Dynamics \textbf{10}%
, 1--18 (1996).

\bibitem{Luo2009a} A. C. J. Luo, Y. Guo, Motion Switching and Chaos of a
Particle in a Generalized Fermi-Acceleration Oscillator, Mathematical
Problems in Engineering, vol. \textbf{2009}, Article ID 298906, 40 pages,
2009. doi:10.1155/2009/298906.

\bibitem{Luo2009b} A.C.J.Luo, \textit{Discontinuous Dynamical System on
Time-varying Domains}. Series: Nonlinear Physical Science. Higher Education
Press, Beijing and Springer, Dordrecht, Heidelberg, London, New York (2009).

\bibitem{Giusepponi2003} S. Giusepponi, F. Marchesoni, The chattering
dynamics of an ideal bouncing ball, Europhysics Letters \textbf{64}, 36
(2003).

\bibitem{Giusepponi2005} S. Giusepponi, F. Marchesoni, M. Borromeo,
Randomness in the bouncing ball dynamics, Physica A \textbf{351}, 142--158
(2005).
\end{thebibliography}
\end{document}